\documentclass[useAMS,usenatbib]{mn2e}

\usepackage{graphicx} 
\usepackage{names} 
\usepackage{times}
\begin{document}

\title{Tracing spiral density waves in M81.}
\author[S. Kendall, R. C. Kennicutt, Jr., C. Clarke, M. D. Thornley]{S. Kendall$^1$\thanks{E-mail: sak39@ast.cam.ac.uk}, R. C. Kennicutt$^1$, C. Clarke$^1$, M. D. Thornley$^2$\\
$^1$ Institute of Astronomy, University of Cambridge, Madingley Road, Cambridge CB3 0HA\\
$^2$ Department of Physics and Astronomy, Bucknell University, Lewisburg, PA 17837}
\date{submitted, accepted}

\pagerange{\pageref{firstpage}--\pageref{lastpage}} 

\maketitle

\label{firstpage}

\begin{abstract}

We use \textit{SPITZER} IRAC 3.6 and 4.5$\mu$m near infrared data from the Spitzer Infrared Nearby Galaxies Survey (SINGS), optical \textit{B, V} and \textit{I} and 2MASS \textit{K$_{s}$} band data to produce mass surface density maps of M81. The IRAC 3.6 and 4.5$\mu$m data, whilst dominated by emission from old stellar populations, is corrected for small-scale contamination by young stars and PAH emission. The \textit{I} band data are used to produce a mass surface density map by a \textit{B-V} colour-correction, following the method of Bell and de Jong. We fit a bulge and exponential disc to each mass map, and subtract these components to reveal the non-axisymmetric mass surface density. From the residual mass maps we are able to extract the amplitude and phase of the density wave, using azimuthal profiles. The response of the gas is observed via dust emission in the 8$\mu$m IRAC band, allowing a comparison between the phase of the stellar density wave and gas shock. The relationship between this angular offset and radius suggests that the spiral structure is reasonably long lived and allows the position of corotation to be determined.

\end{abstract}


\section{Introduction.}\label{intro}

Although many theories have been proposed, the origins and driving forces of spiral structure in galaxies are still not particularly well understood. It is widely accepted that observed grand design spiral arms are tracers of underlying mass variations in galaxies. It is possible that these grand design spirals are long-lived structures, independently able to maintain their shape for many galactic rotations (QSSS - \cite{1964ApJ...140..646L, 1966PNAS...55..229L}), although the existence of isolated long-lived density waves has been questioned, as simulations are only able to reproduce transient spiral structure (e.g. \cite{1984ApJ...282...61S}). If grand design spirals are driven rather than spontaneous, the driving force could be a central bar or external interactions with companion galaxies \cite{1979ApJ...233..539K, 1982MNRAS.201.1035E}. 

It is clear from studying spiral galaxies that star formation can be closely linked with spiral structure, and is preferentially located on spiral arms. It has even been argued that the observed spiral arms are in fact an effect, rather than cause, of the star formation (self-propagating star formation \cite{1976ApJ...210..670M, 1978ApJ...223..129G}; a chain of supernova shocks, or some other star-forming process, induces star formation which is then sheared into spiral arms by the differential rotation of the galaxy). This theory has been disproved for many large grand-design galaxies through the observation of underlying spiral arms in the old stellar population (\cite{1976ApJS...31..313S}, \cite{1989ApJ...343..602E}, \cite{1990ApJ...355...52E}, \cite{1993ApJ...418..123R}), but is still an intriguing possibility for flocculent galaxies, although some flocculent galaxies have also been shown to have weak underlying spiral structure \cite{1999AJ....118.2618E, 1996ApJ...469L..45T}.

Many surveys have been carried out using relatively large samples of spiral galaxies observed in the near infrared, including work by \cite{1984ApJS...54..127E}, \cite{1995ApJ...447...82R}, \cite{1998MNRAS.299..685S}, and \cite{2004A&A...423..849G} which identify underlying density waves and place constraints on the relative amplitude of the stellar arms (as a fraction of the axisymmetric components). The amplitude of the spiral wave is an important input parameter in predicting the response of the gas to the stellar potential. In some cases, these studies also contrast the infrared stellar mass surface density variation with star formation via optical images (\cite{1984ApJS...54..127E, 1998MNRAS.299..685S, 2002MNRAS.337.1113S}). Other morphological studies have looked at the phases of spiral arms - e.g. studies of pitch angle \citep{1942AnLun..10....7D, 1981AJ.....86.1847K, 1982ApJ...253..101K}. Some studies find evidence of angular offsets between the stellar waves and star forming features or dust lanes has been found, e.g.  \cite{1998MNRAS.299..685S} and tentatively \cite{1981AJ.....86.1847K}.

The relative locations of the spiral density maxima (and hence potential minima) and the shock front in the gas can be used as an indicator of the long-term behaviour of the stellar spiral because the response of the gaseous disc to the spiral potential is strongly influenced by the lifetime of the spiral pattern. There are three possibilities; i) a transient spiral where the lifetime is of order the dynamical time, such as generated spontaneously in an isolated galaxy, will induce shocks in the gas that are located on the potential minima (e.g. \cite{1984ApJ...282...61S, 2006MNRAS.371..530C}). ii) If the spiral structure is long-lived, steady-state behaviour has time to emerge (a true density wave in the QSSS interpretation). The steady state response of the gas is to form a shock at an angular offset which lies upstream from the density wave at large radii \citep{1969ApJ...158..123R, 1972ApJ...173..557S, 2004MNRAS.349..909G}. iii) Alternatively, if - as is likely in the case of M81 - the spiral structure is induced by interaction with a companion, the lifetime of the pattern is likely to be intermediate between the cases i) and ii) above. The consequences of this scenario for the relationship between the arms and shocks has not been explored to date. Thus the behavior of the shock indicates whether the spiral pattern is a long-lived or transient feature, and if the former, potentially provides a way of constraining corotation without resorting to arguments based directly on the morphology.

M81 has of course been studied in detail in the past, perhaps most notably by \cite{1980A&A....88..159V, 1980A&A....88..149V} through the use of HI dynamics combined with the density wave amplitude as determined by \cite{1976ApJS...31..313S}. However, despite the extensive past research into spiral structure (and M81 in particular), only a small fraction of which is discussed above, the subject is by no means fully understood. The increased depth and resolution of infrared images provided by the \textit{Spitzer Space Telescope} offers an ideal opportunity to return to the subject. In this paper we produce maps of stellar mass surface density in order to determine the morphology of the stellar spiral structure, including the pitch angle and amplitude of the spiral arms. We also identify the shock front in the gas, and quantify the offset between the peak of the spiral arms and gas shock in order to determine the radius of corotation. 

In addition to the direct application in determining spiral structure, a reliable and simple method of reproducing the underlying mass distribution in a galaxy would be very useful. Potential applications include studies of secular evolution of galaxy discs (through quantifying torques), identifying weak spiral arms in optically flocculent galaxies, and studying the relationship between star formation and spiral arm strength, in addition to providing accurate maps of the potential distribution for input into simulations.

\section{Methods.}\label{methods}

\subsection{Overview of Data.}\label{data}

In order to study stellar spiral structure it is necessary to produce maps of stellar mass surface density. The emission in the near infrared (NIR) is dominated by old populations - largely red giants - and will thus be an excellent tracer of the underlying mass distribution. However, some contamination arises from young stars - red supergiants (RSGs), and OB associations. Previous studies have shown these contributions to the flux to be limited to small spatial regions and even in these regions at most 20 per cent of the flux comes from the 'contaminating' population \citep{1993ApJ...418..123R}. In addition, the effects of interstellar dust are greatly reduced in the near infrared (studies of \textit{K$_{s}$} band (2.2$\mu$m) data have demonstrated that the flux is attenuated by no more than 10 per cent, even in dust lanes \citep{1993ApJ...418..123R}). The longer wavelengths detected by IRAC are likely to suffer even less attenuation. Despite the suitability of the near-infrared for this type of work, due to the inherent uncertainties outlined above it was decided to use several complimentary approaches to allow cross checks in our methodology. The primary data used for this research was obtained on the \textit{Spitzer Space Telescope} by the SINGS project \citep{2003PASP..115..928K, 2004ApJS..154..222W}. IRAC bands 1, 2 and 4 (3.6, 4.5 and 8$\mu$m) are used; the 3.6 or 4.5$\mu$m can be thought of as tracers of the underlying stellar mass distribution. Band 4, the 8 $\mu$m data, is used as a tracer of the shocks induced in the gas (further discussion in section \ref{accuracy}). The IRAC data reduction is described in \cite{2004ApJS..154..204R} and the SINGS documentation. The IRAC data have a pixel scale (after drizzling) of 0.75 arcsec, and sensitivities 25.7 and 23.0 mag(AB) arcsec$^{-2}$ (1$\sigma$) for 3.6 and 4.6$\mu$m respectively. The PSFs are approximately 1.7 arcsec for 3.6 and 4.5$\mu$m, and 2 arcsec for the 8$\mu$m data. In addition, optical images in \textit{B}, \textit{V}, and \textit{I} bands were used to create a mass map of the galaxy using a colour-correction technique described further in section \ref{BVRI}. The optical images were obtained on the University of Arizona's Bok Telescope using the 90prime instrument \citep{2004SPIE.5492..787W}. Details on the data reduction can be found in \cite{2006ApJ...648..987P}. The pixel scale of the \textit{B}, \textit{V}, and \textit{I} bands is 0.45 arcsec pix$^{-1}$, with sensitivities of 26.7, 26.3 and 27.9  mag(AB) arcsec$^{-2}$ respectively. The PSFs are approximately 1.8 arcsec for all three bands. Finally, \textit{K$_{s}$} band (2.2$\mu$m) images from the 2MASS catalogue \citep{2000AJ....119.2498J} are used. The \textit{K$_{s}$} band images have a sensitivity of 18.5 mag(AB) arcsec$^{-2}$ and a pixel scale of 1.00 arcsec pix$^{-1}$ (PSF $\sim$3 arcsec). One arcsecond corresponds to $\sim$17pc at 3.6Mpc, so at a radius of 5kpc from the galactic centre, even with the largest scale images one pixel is equivalent to 0.1$^o$. When compared with the error estimates for azimuthal angle quoted in section \ref{accuracy} it can be seen that the pixel scale is not a limiting factor in this analysis.\\

\subsection{IRAC 3.6 and 4.5$\mu$m data.}\label{IRAC}

To extract the stellar spiral structure from the 3.6$\mu$m data the axisymmetric components of the galaxy were first removed from the images. The non-axisymmetric components (residuals) can then be corrected for non-negligible small scale contamination by foreground stars and dust emission (features up to $\sim$10 arcsec in diameter). The order of these steps is initially necessary because the correct adjustments require a visual inspection of the residuals (and so an axisymmetric model must be subtracted first). However an iterative process would be possible, and in some cases preferable, in which the final fit to the axisymmetric components is carried out after the small-scale contamination has been removed. In the case of M81 the difference is negligible because the star formation rate is relatively low, and the small-scale non-axisymmetric features make little difference to the fitting of the axisymmetric components. Although a stellar mass surface density map is not directly produced via the process described below, one can be easily created through the combination of the axisymmetric model with the fully cleaned residual images. 
 
To fit the axisymmetric components, GALFIT \citep{2002AJ....124..266P} was used to produce a three-component fit to the galaxy using a 2d implementation of a Sersic bulge (with index n=2.62 and R$_{e}$=46.2 arcsec), exponential disc (R$_{s}$=155.4 arcsec) and constant background. The axis ratios (b/a) and position angles determined for the bulge were 0.71 and -31.9$^\circ$ and for the disc were 0.52 and -28.3$^\circ$ respectively. GALFIT has the ability to produce a model galaxy based on the best-fitting parameters, and this model was subtracted from the 3.6$\mu$m image to reveal the non-axisymmetric components, dominated by the spiral arms. The GALFIT model is discussed further in section \ref{accuracy}.

PAHs (polycyclic aromatic hydrocarbons) have emission peaks, in addition to continuum emission. A PAH emission feature at 3.3$\mu$m (\citep{1991ApJ...380..452T, 1981MNRAS.196..269D}) falls within the IRAC 3.6$\mu$m bandwidth, and so this is the most likely cause of much of the contamination in the residual image. PAH emission is also found in the 8$\mu$m waveband, and this was used to reduce the contamination in the 3.6$\mu$m data by subtracting a scaled version of the 8$\mu$m data. The exact method is as follows; the 8$\mu$m data were corrected for the stellar continuum emission by the subtraction of a scaled version of the 3.6$\mu$m GALFIT model galaxy. The scaling constant used was 0.232, as given by \cite{2004ApJS..154..253H}. It should be noted that, in using the GALFIT model (rather than the original 3.6$\mu$m image) to subtract the stellar continuum from the 8$\mu$m data, the continuum contribution from the spiral arms is not taken into account. However, the original image has been shown to contain PAH emission; by using the GALFIT model we remove the risk of affecting the PAH contribution to the 8$\mu$m emission. The effect of this approximation (after all corrections), as can be seen in equation \ref{eq1}, is that the 3.6$\mu$m flux will be slightly lower on the spiral arms than if the full 3.6$\mu$m image been used to remove the continuum, but by less than a factor of $\sim$0.05. This small change is unlikely to be noticeable above the noise in the data. The systematic effect on the phase (if noticeable) will be to reduce the offset measured between the 8$\mu$m peaks and the density maximum in the stellar spiral wave by slightly increasing the amplitude of the 8$\mu$m feature in phase with the stellar spiral.

After correcting for the stellar continuum the 8$\mu$m image should have no remaining axisymmetric components. This continuum-corrected 8$\mu$m image can then be multiplied by a scaling constant and subtracted from the non-axisymmetric 3.6$\mu$m residuals to remove the PAH features. The scaling constant used was 0.08, chosen (by eye) to maximize the smoothness of the arms after the PAH subtraction, and this is judged to be accurate to $\sim\pm$5 per cent. The process can be described by the following equation;\\

\begin{equation}
  {3.6_{corr.} = 3.6_{non-axisym.}-0.08(8_{data}-0.232(3.6_{model}))}
  \label{eq1}
\end{equation}

The justification for the subtraction of PAH components from the 3.6$\mu$m image is relatively straightforward; a visual inspection of the data before and after the PAH features are removed shows a dramatic improvement in the smoothness of the arms (and any underlying spiral structure will produce a smooth mass distribution). The scaling constant used is in fact consistent with values suggested by models of PAH emission and observations of interstellar emission the Galaxy (see, e.g. \cite{2007ApJ...657..810D, 2006A&A...453..969F}): \cite{2006A&A...453..969F} find a range in the ratio of 3.6/8$\mu$m intensities from 0.059-0.094 over 6 different regions observed, with an average value of 0.072. 

It is likely that the flux from star forming regions also has small contributions from OB associations and RSG stars as discussed in section \ref{data}. In addition, a number of foreground stars overlay the galaxy. The removal of remaining fine-structure was achieved by using the IRAF task XZAP, which removes any features which are n$\sigma$ above the local background. The size of the features removed, and value of n, can be adjusted to give the most appropriate corrections. XZAP was slightly modified to use MEDIAN rather than FMEDIAN when calculating the smoothed image; the quantization of the data by FMEDIAN was found to have a noticeable effect in the fainter regions of the residual image. The best-fitting corrections turned out to have only a weak dependence on $\sigma$, but the size of the smoothing box used was critical - too small and the small-scale features would be left virtually unchanged, whereas if the smoothing box was too large the features of interest could be affected. The best values were chosen after careful examination by eye of the final images in comparison to the original data. Final editing by hand was required for the largest foreground stars in the image, and this was carried out using IMEDIT in IRAF. Although the removal of stars in this way potentially leaves residual features from the PSF wings, these will be small in comparison to the PAH emission regions, and varying on scales that are much smaller than the spiral pattern.

\begin{figure*}   \centering
  \vbox to200mm{\vfil 
    \includegraphics[width = 180mm]{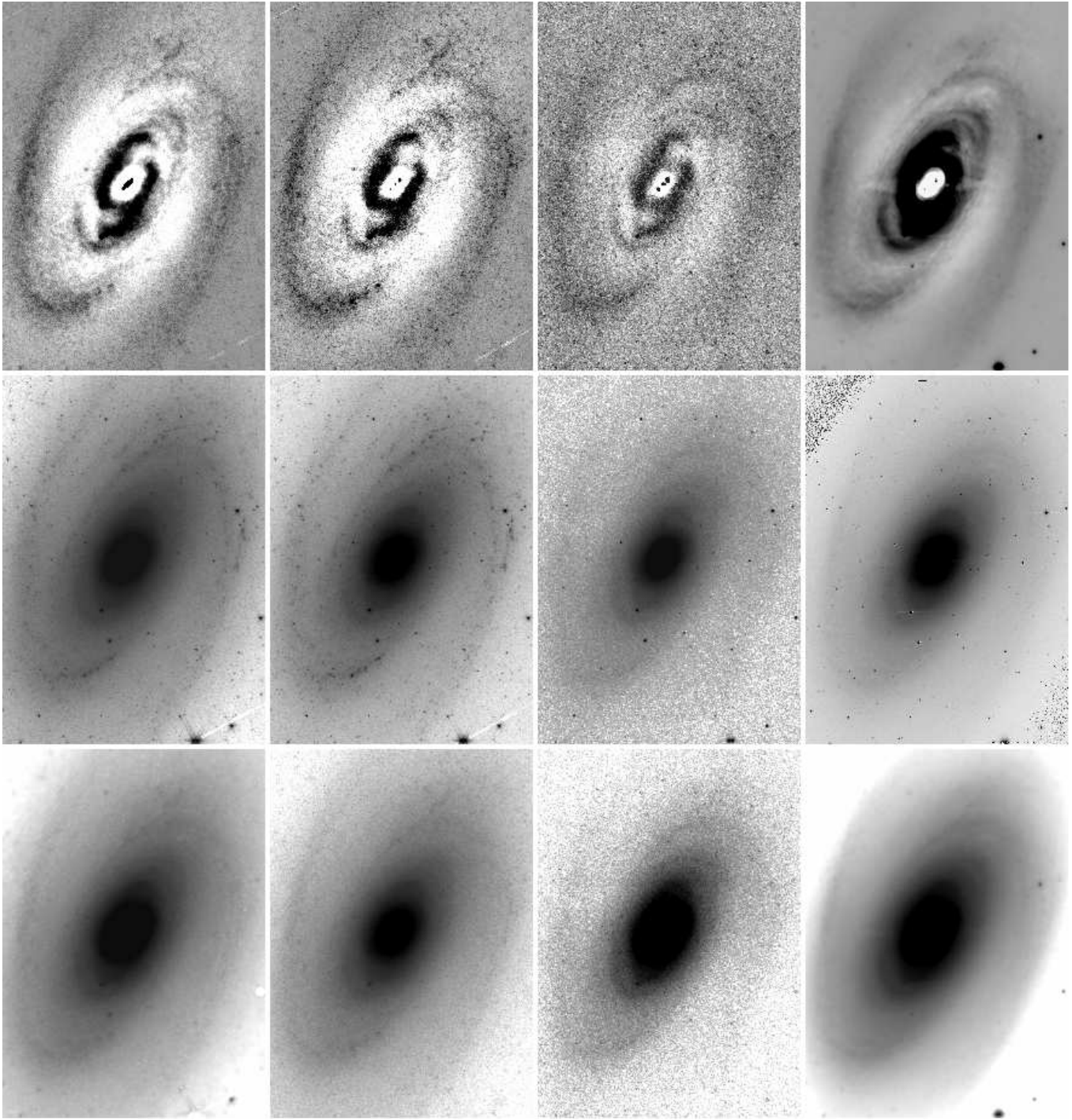}
  \caption{The non-axisymmetric residuals for (top, l-r) IRAC 3.6 and 4.5$\mu$m, \textit{K$_{s}$} band, and \textit{I} band colour-corrected mass maps respectively. The middle row shows the original data for the IRAC 3.6 and 4.6$\mu$m, \textit{K$_{s}$} band, and the colour-corrected \textit{I} band before the axisymmetric component is removed. The bottom row shows the mass surface density maps for each waveband in turn.}
  \label{figrelamp}
 \vfil}
\end{figure*}

The IRAC 4.5$\mu$m image was also used, to provide a comparison with the 3.6$\mu$m data. The 4.5$\mu$m data also appear to have PAH contamination, although with a smaller correction needed (a factor of 0.05 of the 8micron continuum-corrected image, as compared with 0.08 for 3.6$\mu$m. This is consistent with measured ratios of 4.5/8$\mu$m flux in \cite{2006A&A...453..969F} who find ratios in the range 0.037-0.065, with an average of 0.048. Other steps in the image processing were exactly as for 3.6$\mu$m and will not be described further.

The process of creating a mass map can be seen in Fig. \ref{figrelamp}; the top row shows the residual non-axisymmetric components after removing the PAH components (for the IRAC data) and running XZAP. These residual images are combined with the GALFIT models (not shown) to create the mass surface density maps on the bottom row. the original data is shown  for comparison.

\subsection{Optical data.}\label{BVRI}

Optical \textit{B},\textit{V} and \textit{I} band images were used to create a stellar mass surface density map of the galaxy. The method relies on the relationship between mass-to-light ratio (M/L) and optical colour as described by \cite{2001ApJ...550..212B} (BdJ) using \textit{B-V} pixel-to-pixel colour to correct an \textit{I} band luminosity map of the galaxy, using the relationship log$_{10}$(M/L$_I$) = a$_I$ + b$_I$(\textit{B-V}). In their 2001 paper, BdJ give values for a$_I$ and b$_I$ as found from galaxy evolution models (the zeropoint, a$_I$, assumes maximum disc M/L ratio). In this paper, as with PAH corrections for the IRAC data, the best-fitting value of b$_I$ is judged by eye so as to maximize the smoothness of the mass distribution. The value of b$_I$ quoted by BdJ is higher than appears ideal in this case (see Table \ref{tab1}). The relative amplitude of the spiral structure (to axisymmetric components) turns out to be relatively insensitive to the value of b$_I$ used; almost doubling the value of b$_I$ from 0.275 to 0.5 only reduces the relative amplitude of the spiral arms by an average of $\sim$15 per cent, which is less than the discrepancies between relative amplitude estimates from the different wavelengths, as will be shown in section \ref{amp}. In contrast, the value of a$_I$ only affects the zeropoint of the mass output, which is not important when considering the relative amplitude of the spiral arms and so this was not investigated further. As a result, readers are advised to exercise caution if creating normalised mass-models using the values of a$_I$ and b$_I$ found in this work.

\begin{table}
  \centering
  \begin{tabular}{|l|l|l|} 
    \hline 
    Method & a$_I$ & b$_I$ \\ 
    \hline
    Bell \& de Jong 2001 & -0.627 & 1.075 \\
    Bell et al 2003 & -0.399 & 0.824 \\
    Visual examination of these data & -0.627 & 0.275 \\ 
    \hline
  \end {tabular}
  \caption{The comparison of the best-fitting values of the  constants a$_I$ and  b$_I$ as used in the method log$_{10}$(M) = log$_{10}$(L$_I$) + a$_I$ + b$_I$(B-V). The value of a$_I$ affects the normalisation of the stellar mass into solar units, and depends on whether the disc is sub-maximal or not. For this paper the value of -0.627 was chosen to be consistent with \protect\cite{2001ApJ...550..212B}.}
  \label{tab1}
\end{table}

In a later paper, \cite{2003ApJS..149..289B} offer a potential explanation for the discrepancy demonstrated in Table \ref{tab1}, with the acknowledgment that the original models do not use as large a metallicity scatter as observed in real galaxies. This metallicity scatter may serve to over-estimate the M/L slope, particularly in the NIR, although the corrections quoted do not reduce the M/L ratio in the \textit{I} band sufficiently (the ratio decreases by a factor of $\sim$1.3 rather than $\sim$3 as needed to reconcile the differences in Table \ref{tab1}). However, given the large scatter in the data (fig 20. \cite{2003ApJS..149..289B}), the lower M/L ratio found in this case would appear to lie within reasonable limits.

The rationale for the choice of \textit{B}, \textit{V} and \textit{I} bands is partly necessity, since data were available in \textit{B}, \textit{V}, \textit{R}, and \textit{I} bands only. However, BdJ recommend using an optical colour such as \textit{B-V} or \textit{B-R}, because these are effectively tracers of age (rather than metallicity as would be the case with a NIR colour). BdJ further note that a NIR band is best for mass estimates because the \textit{K$_{s}$} or \textit{I} bands have much smaller dynamic ranges than their optical counterparts.

Once the mass map of the galaxy has been produced, the subsequent steps are as described above for the IRAC data; GALFIT is used to remove the axisymmetric component of the galaxy, and XZAP for star and fine-structure removal. \\

\subsection{2MASS data.}\label{2MASS}

The analysis of the 2MASS \textit{K$_{s}$} band data was rather simpler than the previous two methods described above, possibly due to the relatively poor resolution and depth of the data (18.5 (AB)mag arcsec$^{-2}$ for \textit{K$_{s}$} compared to 25.7 (AB)mag arcsec$^{-2}$ for 3.6$\mu$m). The two steps required were the fitting of an axisymmetric component by GALFIT and then XZAP to remove foreground stars and fine structure. 

\subsection{IRAC 8$\mu$m data.}\label{8mic}

\begin{figure}   \centering
  \includegraphics[width=83mm]{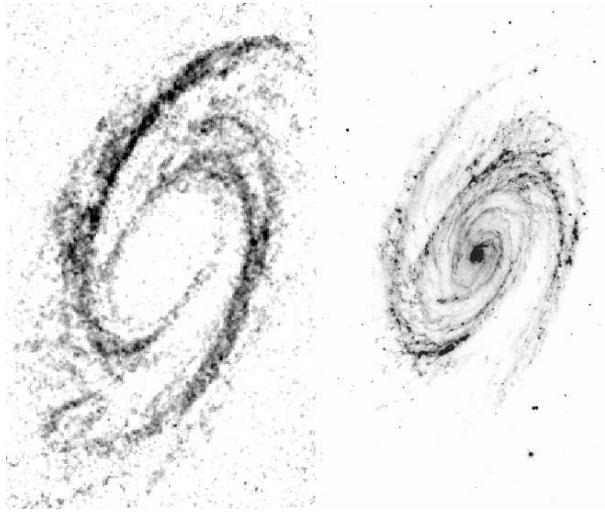}
  \caption{HI observation (known to trace shocked gas) on left, \textit{Spitzer} 8$\mu$m on right, showing the degree of agreement between the emission at the two wavelengths. Images are to the same scale. The HI observation was obtained from NED, originally published by \protect\cite{1995A&AS..114..409B}.}
  \label{figHIcomp}
\end{figure}

The gas shock can be traced through the 8$\mu$m because this wavelength largely traces dust emission. A more conventional tracer of gas shocks is the HI 21cm line, but the 8$\mu$m band has several advantages because any galaxy in the SINGS sample has high resolution 8$\mu$m data available. Emission at 8$\mu$m is dominated by dust, which tends to be concentrated in regions of high gas density, which occurs at (or just behind) the shock front. To be visible in emission the dust needs to be heated, and the primary mechanism for concentrated emission is the switch-on of young stars triggered by the shock front. The link between the two wavelengths is highly plausible (see Fig. \ref{figHIcomp}), although not all 8$\mu$m emission is from HI regions. The 8$\mu$m image is prepared for use by subtracting the stellar continuum emission, as described in section \ref{IRAC}, but is otherwise unchanged.

\section{Discussion of mass surface density estimates.}\label{accuracy}

The two main quantities derived from the residual mass surface density maps are the phase and amplitude of the stellar spiral, and as such it is important to have a good grasp of the uncertainties in measuring these quantities. 

Over- or under-correction for PAH contamination via the scaled 8$\mu$m image will lead to errors in the flux, and hence mass surface density. This is most likely to affect areas near the peak of the mass surface density because dust is concentrated in the shock. It is hard to quantify the possible error in the relative amplitude as a result of remnant PAH emission after all corrections, but it is possible to put a upper bound on the problem; a comparison of the relative amplitude for fully-corrected data with the measured relative amplitude if no PAH or fine structure correction is used gives a difference of no more than $\sim$20 per cent, and at most radii no more than 15 per cent. The remaining error is likely to be a small fraction of this. The equivalent step for the optical images, colour-correcting the \textit{I} band mass map will lead to an incorrect mass/light ratio and hence incorrect mass estimate. Dust lanes, star-forming regions and other features that are prominent in the \textit{B-V} map will be particularly sensitive to the choice of M/L ratio, and as such this could introduce a systematic error in the peaks of the spiral arms and an incorrect value for the amplitude of the spiral. A scaling constant that is too large will over-subtract from the star forming regions, and increase the intensity in the dust lanes (the opposite is true if the scaling constant is under-estimated). In addition, the choice of smoothing box size in XZAP must be carefully considered; if the smoothing box is made too large this has the potential to remove features that are part of the wave which would lead to systematic under-estimates of the amplitude of the spiral arms. Investigation into the effects of the choice of sky sigma and (pixel) width of radial samples in ELLIPSE demonstrates that the errors in relative amplitude are no larger than $\sim$5 per cent for reasonable values. GALFIT was constrained to keep the same ellipticity and position angle (e and PA) values for all fits to all wavelengths, but scalelengths were initially allowed to vary. Fitted this way, the scalelengths for the two IRAC bands varied by $<$1 per cent. In contrast, the fit to the \textit{I} band colour-corrected image gave a disc scalelength that was $\sim$15 per cent smaller. When the fit is constrained to take the same scalelengths as the IRAC 3.6$\mu$m fit, the relative amplitude of the spiral structure varies by no more than 5 per cent compared to the unconstrained fit over most radii, although the difference is as large as $\sim$15 per cent around R = 300arcsec.

Examination of alternative fits to the PA showed that the sensitivity is $\sim\pm$3$^o$. Within this range the residual image is not noticeably affected, but the fit quickly appears worse as the PA is varied by larger amounts. As a further confirmation that the values used in this work are reliable, the PA of the exponential disc fitted by GALFIT agrees well with position angles determined kinematically (e.g., within one sigma of the value found in \cite{1975A&A....45...43R} and within 2 degrees of \cite{1980A&A....88..149V}). The fit in e is similarly good. As with the amplitude, constraining the \textit{I} band scalelengths during fitting with GALFIT caused minimal changes to the phase; the fit to the gradient of the log spiral differed by no more than $\sim$2 per cent over the region in which it follows a log spiral. 

As previously described, the PAH correction factor affects the mass distribution near the peaks of the spiral arms, which can create artificial trends in phase if the PAH correction is not correct. Depending on the method used the phase can be quite insensitive to PAH contamination because these peaks are much sharper than the underlying spiral. Using sine waves to fit the phase produces errors of no more than 7$^o$ between phase measured for fully corrected residuals and those with no PAH or small-scale structure corrections. The errors are not random, tending to steepen the gradient of the log spiral, thus reducing the pitch angle (by a maximum of $\sim$5 per cent in gradient). However, if instead a single point of highest intensity is used to measure the phase of the stellar spiral the error between phase estimates at each radius can be as much as $\sim$60$^{o}$ (discussed further in section \ref{massphase}). The \textit{I} band colour correction has the potential to introduce systematic errors in phase because the offset of the dust lanes and stellar spiral varies with radius. However, provided a suitable choice of M/L ratio is used this will not be a problem. The remaining small scale structures, mostly stars, are unlikely to bias the phase estimates. A small, well defined peak in intensity from a star is clearly distinct from the slow variation in underlying spiral structure. This is especially true where sine waves are being used to estimate the phase of the wave (see section \ref{Results}), but even the methods of determining phase which are most susceptible to sharp peaks in intensity cannot introduce systematic errors into the results, since the positions of foreground stars are random.

Finally, it is worth noting that if the 3.6 and 4.5$\mu$m images are compared, in addition to the large-scale similarities, the majority of small-scale (few pixel) structure is duplicated in both bands. The features observed are most likely individual AGB stars, meaning that much of the small-scale 'noise' is due to these stars. Since AGB stars bias the mass-light ratio, these stars will be the cause of much of the small-scale noise in the azimuthal profiles (see section \ref{Results}).

\section{Results.}\label{Results}

Given residual mass-maps with well defined errors it becomes possible to analyse the data for density wave features. The IRAF task ELLIPSE was used to fit isophotal ellipses to a galaxy profile, and can then extract intensity profiles as a function of azimuth. The axisymmetric models of the galaxy produced by GALFIT were used to allow ELLIPSE to fit the position angle and ellipticity (PA and e) at different radial values (measured on the semi-major axis (SMA)). The e and PA values were then used to make elliptical intensity profiles of the residual images. Ellipses were initially sampled every 5 pixels along the SMA between 250-1000 pixels, with each profile averaging over the 5 pixels width to make use of all available data. These were then combined to create approximately logarithmic steps, such that steps between successive profiles become 5, 10, 15 or 20 pixels as radius increases. Finally, each profile was averaged azimuthally to give angular resolution $\sim$1$^{\circ}$. A sine wave of the form $y = asin(2x+b) + c$ was then fit to each profile in order to extract information about the amplitude and phase of the stellar spiral structure. In addition to elliptical profiles of the galaxy, radial intensity profiles were obtained by sampling the residuals along narrow wedges (4$^\circ$ in angular size) from the centre outwards.

\subsection{Overview of trends in data.}\label{overview}

\begin{figure}   \centering
  \includegraphics[width=83mm]{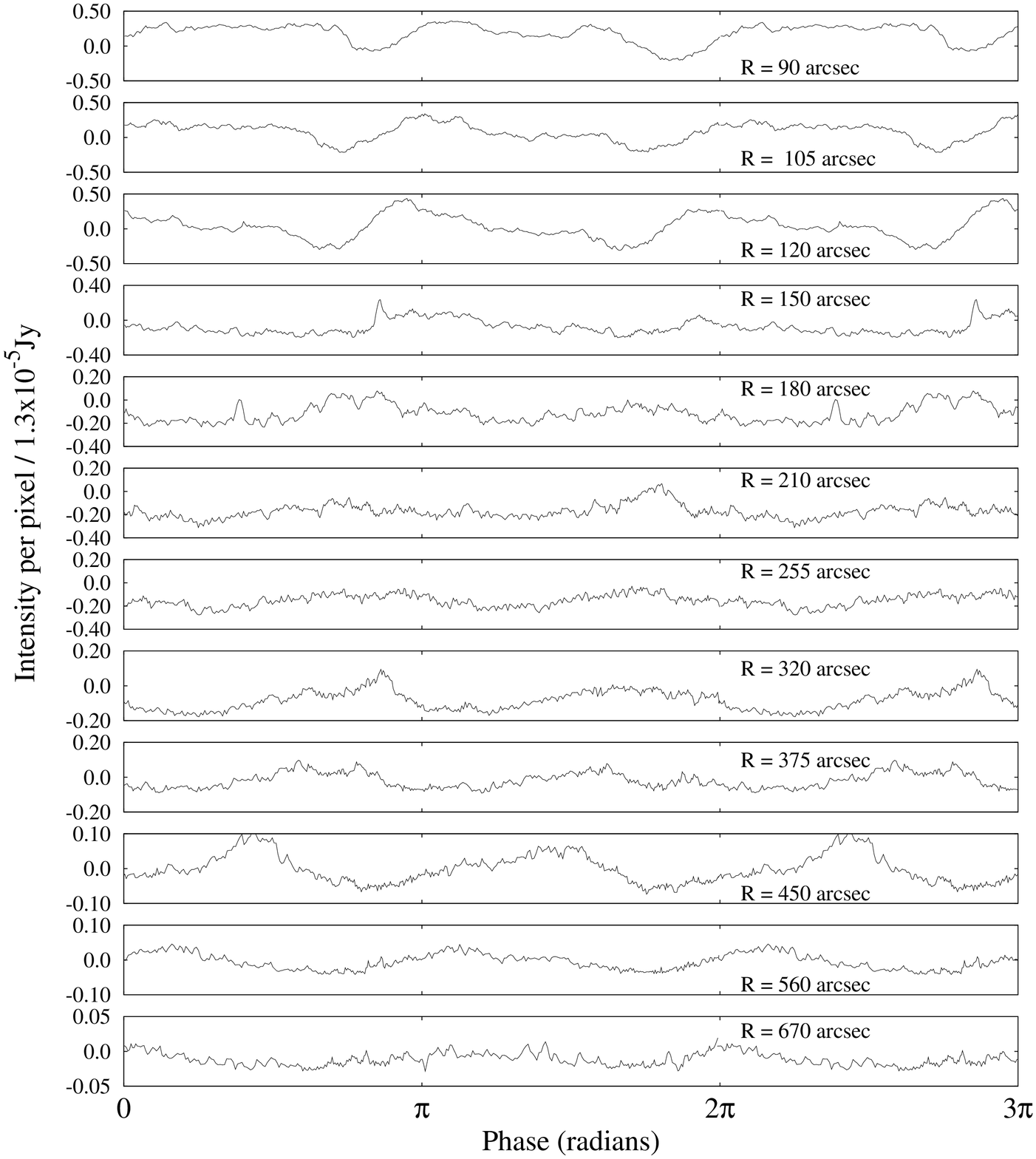}
  \caption{Azimuthal profiles taken from the 3.6$\mu$m residuals. The range in azimuth is extended over 3$\pi$. Zero phase is along the +y axis in Fig. \ref{figazi2}, increasing anticlockwise. The position angle of the disc is -28.3$^{o}$.} 
  \label{figazi1}
\end{figure}

\begin{figure}   \centering
  \includegraphics[width=83mm]{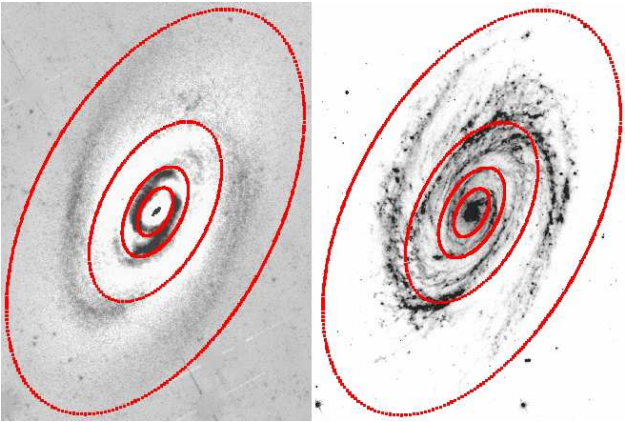}
  \caption{IRAC 3.6 and 8$\mu$m maps of non axisymmetric mass surface density (residuals). Ellipses at 80, 150 300 and 675 arcseconds are marked.}
  \label{figazi2}
\end{figure}

As can be seen in Figs. \ref{figazi1} and \ref{figazi2}, the residuals are dominated by the central ring feature around 100 arcsec. The azimuthal profiles over the range 90$<$R$<$150 arcsec in Fig. \ref{figazi1} show approximately constant intensity except at the small breaks at either end of the ring. The profile at 150 arcsec also starts to detect the features that appear to be short spiral arms (which can be seen in Fig. \ref{figazi2}). The ring has been noted in previous work on M81 \citep{1991SvAL...17..168G,1995AJ....110.2102E}, but the two spiral arms that appear to connect to the ring have not been mentioned previously. The possibility that the inner spiral arms are purely an artefact due to subtracting the axisymmetric model (particularly errors due to the bulge fit) has been considered, but this does not seem to be the case: at 150 arcsec the bulge contributes no more than a quarter of the light from the total axisymmetric components, and the ring and inner spiral can still be identified even if the bulge components are left in the residuals (and the disc alone is subtracted). The features are also immune to adjustments of a few degrees in the disc PA fits, beyond which the entire disc fit is obviously incorrect. In addition, there is sufficient agreement between the 3.6$\mu$m residuals and the structure in the 8$\mu$m (see, e.g. Fig. \ref{figazi2}) to further support the identification of the inner spirals as real features. 

Between $\sim$150 and 300 arcsec the spiral arms appear to vanish: although small perturbations with two-fold symmetry are visible in this range the phase does not vary with radius; maxima are at $\sim$130$^\circ$ - not aligned with the SMA (PA = 152$^\circ$). The amplitude of the perturbations is small compared to the spiral arms beyond 300 arcsec, with the relative amplitude increasing from 0.05-0.1. The cause of the variation in intensity is not clear, but does not appear to be a density wave.

Beyond $\sim$300 arcsec the spiral arms are prominent in the image, and the azimuthal profiles show clear sinusoidal oscillations with a phase dependence on radius. Thus the early indications suggest that a density wave is present in this radial range, and extends unbroken through almost 180$^{o}$. 

In the 8$\mu$m image, Fig. \ref{figazi2}, there is clear evidence of spiral structure. The grand-design spiral arms dominate outside $\sim$300 arcsec and can be traced over 180$^{o}$ of rotation. Inside this radius the structure becomes more filamentary and two-fold symmetry is less well defined. Relatively extended features are still visible, but seldom seem to extend over more than $\sim$90$^{o}$. In addition to the large spiral patterns, there is a great deal of fine structure (spurs and feathering) visible in the 8$\mu$m image. These features can be explained through the shearing of small-scale structures by differential rotation (e.g. \cite{2004MNRAS.349..270W, 2006MNRAS.367..873D, 2006ApJ...646..213K}).

\begin{figure}   \centering
  \includegraphics[width=83mm]{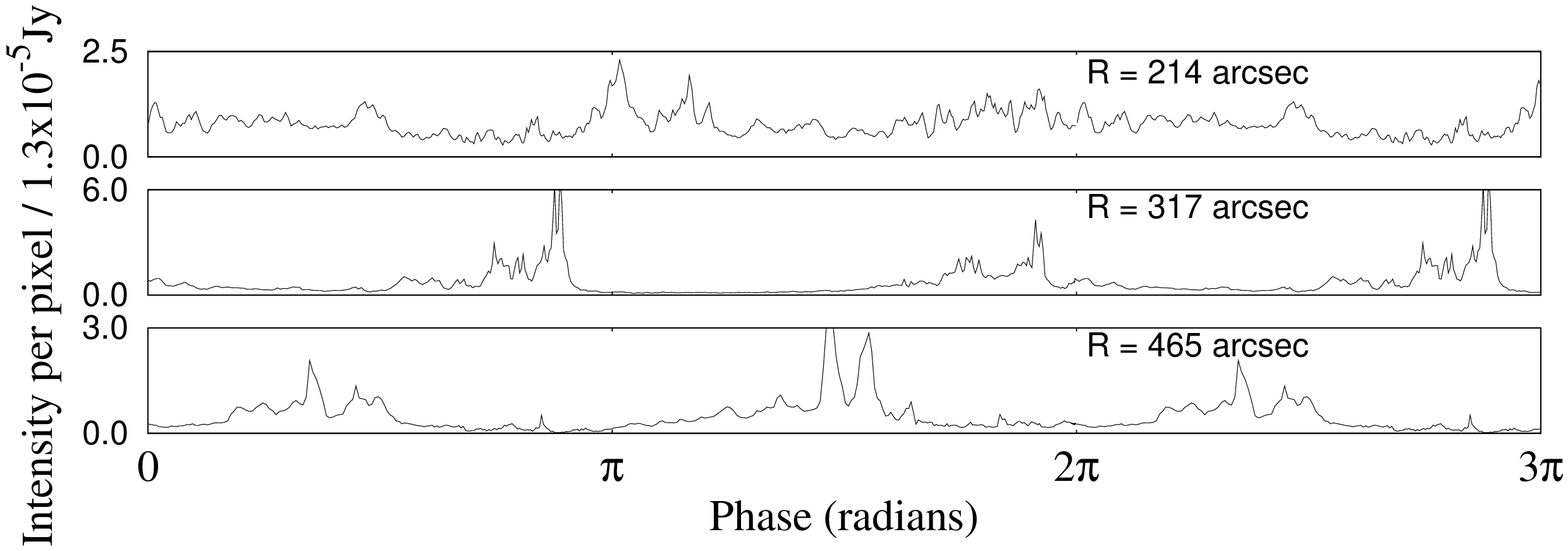}
  \caption{Azimuthal profiles taken from the 8$\mu$m image. The range in azimuth is extended over 3$\pi$. Phase is defined as in Fig. \ref{figazi1}.}
  \label{figazi4}
\end{figure}

\begin{figure}   \centering
  \includegraphics[width=83mm]{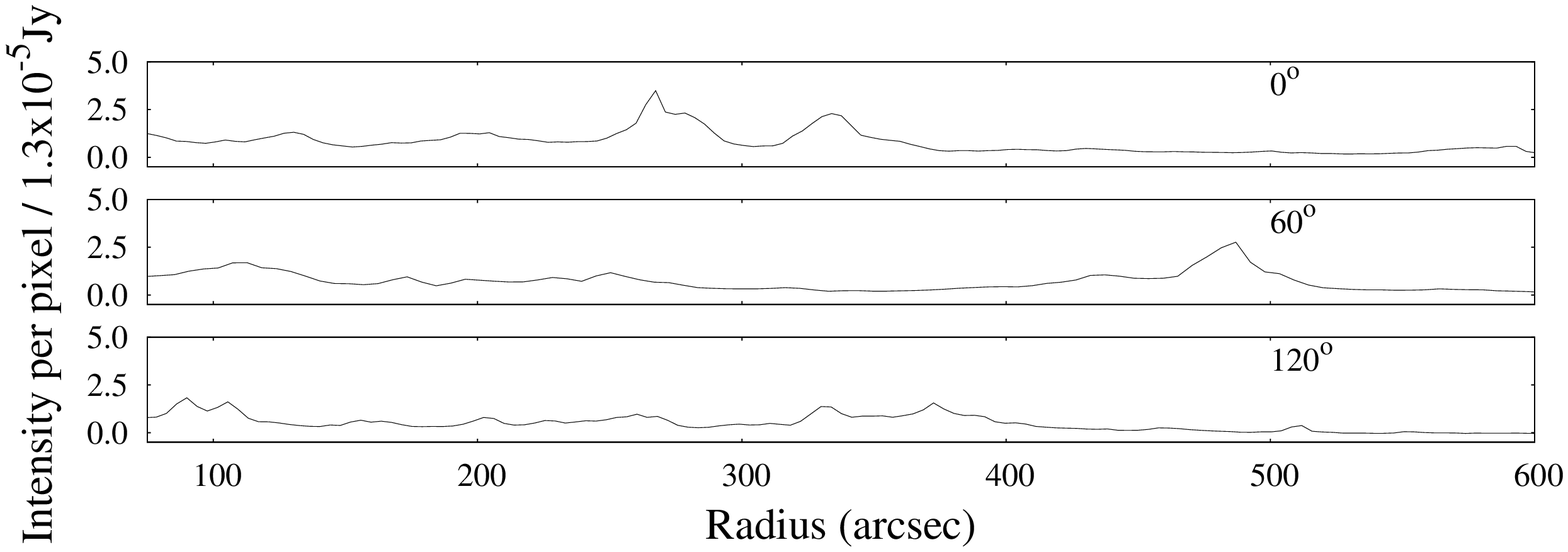}
  \caption{Radial profiles of the 8$\mu$m image. The radius has been corrected to face-on orientation in each case.}
  \label{figrad1}
\end{figure}

There are some features that clearly show up in both the 3.6$\mu$m and 8$\mu$m wavelengths; the well defined spiral arms outside 300arcsec are one such example, as are the much shorter spiral arms around 150arcsec. There is a great deal of fine structure observed at 8$\mu$m that is not found in the stellar component. In the region between 150 and 300 arcseconds it has been noted that, while the 8$\mu$m emission displays spiral-type features, there is little or no evidence for equivalent structure in the stellar distribution. The disjointed nature of the 8$\mu$m emission in this region may be a reflection of the fact that there is no strong stellar wave present. There is also no equivalent in the stellar mass distribution to the spurs and feathering observed in the gas (this is unsurprising given that the mass distribution is expected to be smooth over large scales).

It is striking, although unsurprising, that the 8$\mu$m spiral features are much narrower than the corresponding features in the mass surface density maps, and are far less smooth when viewed along azimuthal profiles (see Figs. \ref{figazi1} and \ref{figazi4}). As will be shown later (section \ref{massphase}) a sinusoidal wave can be used as a good approximation to a density wave but this is clearly not the case for the gas response.

Azimuthal profiles are more effective than radial profiles for studying the morphology of the stellar spiral structure; it is particularly hard to determine the amplitude of the wave from radial profiles because the wave amplitude changes with radius. In contrast, radial profiles are much more accurate for determining the position of the gas shocks from the 8$\mu$m data because the spiral is relatively tightly wound, and so the features are narrow when crossed radially, as can be seen by comparing Figs. \ref{figazi4} and \ref{figrad1}.

\subsection{Amplitudes.}\label{amp}

\begin{figure}  \centering
  \includegraphics[width=83mm]{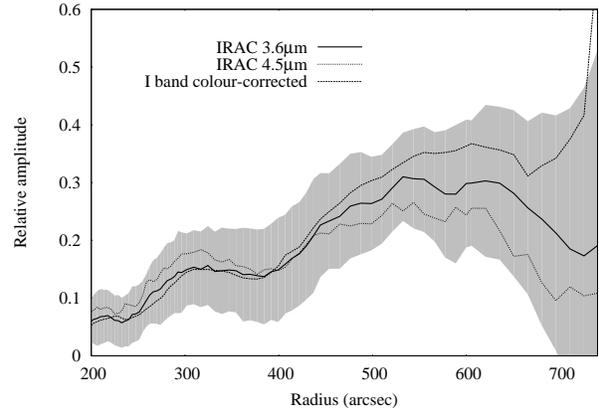}
  \caption{IRAC 3.6$\mu$m, 4.5$\mu$m, and \textit{I} band colour-corrected data showing the relative amplitude of the stellar spiral wave. The shaded area shows the error bars for the 3.6$\mu$m data which are derived from the fits to the azimuthal profiles and are highly conservative. (Error bars for the 4.5$\mu$m, and \textit{I} band data are not shown, but are of a similar scale.}
  \label{figamp1}
\end{figure}

\begin{figure}  \centering
  \includegraphics[width=83mm]{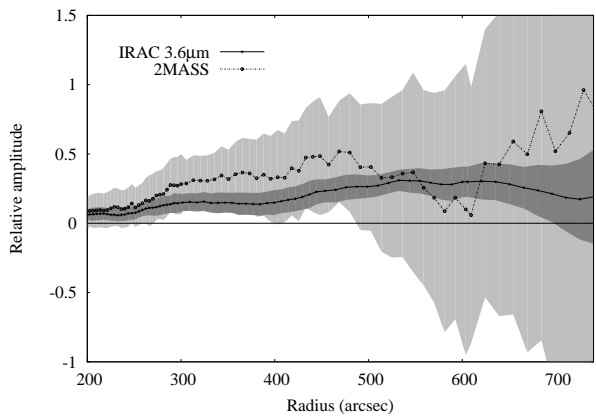}
  \caption{IRAC 3.6$\mu$m and \textit{K$_{s}$} band data for the relative amplitude of the stellar spiral wave showing the sizes of the errors associated with each dataset.}
  \label{figamp3}
\end{figure}

The radial dependence of the amplitude of the stellar spiral structure is shown in Fig. \ref{figamp1}. The results show that both the IRAC and colour-corrected \textit{I} band data display very similar trends of amplitude increasing with radius. The relative amplitude of the 2MASS data is probably suffering from noise effects; examining Fig. \ref{figamp3} shows that the errors in the radial range 300-600 arcsec are much greater for the 2MASS data than for the IRAC data. For this reason it was decided not to include the \textit{K$_{s}$} band data in further analysis. In all data sets the relative amplitude is defined as (a/(model intensity + c)) where a and c are the wave (half) amplitude and average intensity of the azimuthal profile respectively (as defined for the sine wave fit in section \ref{Results}). The denominator is defined to correct for a non-perfect disc fit: whereas a perfect model would have averaged the disc intensity exactly to zero for all radii this has not happened (probably because the galaxy does not have a perfect exponential disc, or as a result of the spiral arms influencing the fit).

\subsection{ Phase of the stellar spiral structure.}\label{massphase}

\begin{figure}   \centering
  \includegraphics[width=83mm]{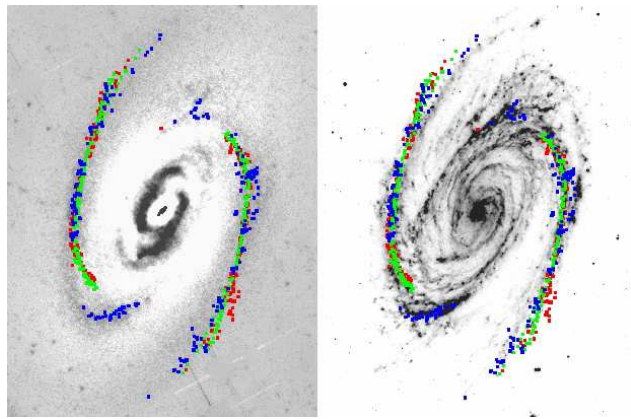}
  \caption{Estimates of the phase of the stellar spiral structure from peak finding (blue) sine wave fits (green) and estimates by eye (red) shown overplotted on the 3.6$\mu$m and 8$\mu$m images. It can be seen that the peak finding method detects contamination from spurs which are clearly visible at 8$\mu$m.}
  \label{figphase2}
\end{figure}

\begin{figure*}  \centering
  \vbox to120mm{\vfil 
  \includegraphics[width=140mm]{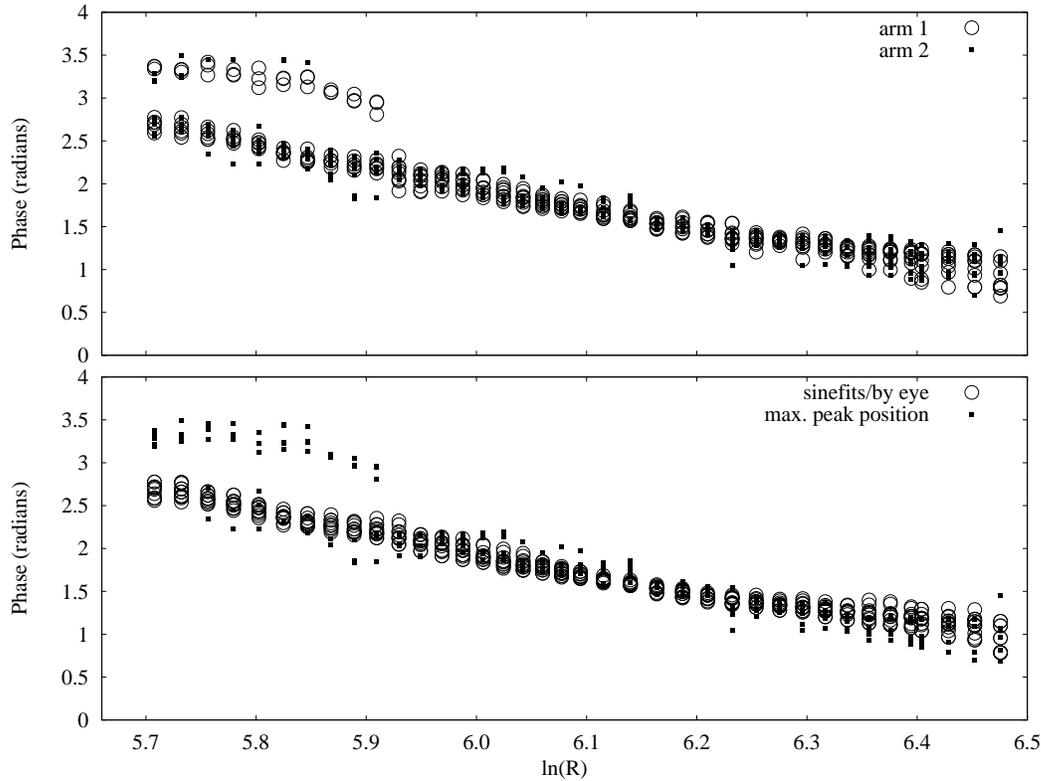}
  \caption{The top plot shows the two arms of the spiral wave with the $\pi$ phase difference removed, demonstrating that although there is scatter, there is no systematic variation between the phases of the stellar arms (they are 180$^\circ$ apart). The bottom plot distinguishes the sine wave fits and fits by eye from the peak positions found from maximum intensity, where it can be seen that there is much more scatter in the latter method. Zero phase lies along the SMA.}
  \label{figphase4}
  \vfil}
\end{figure*}

\begin{figure}  \centering
  \includegraphics[width=83mm]{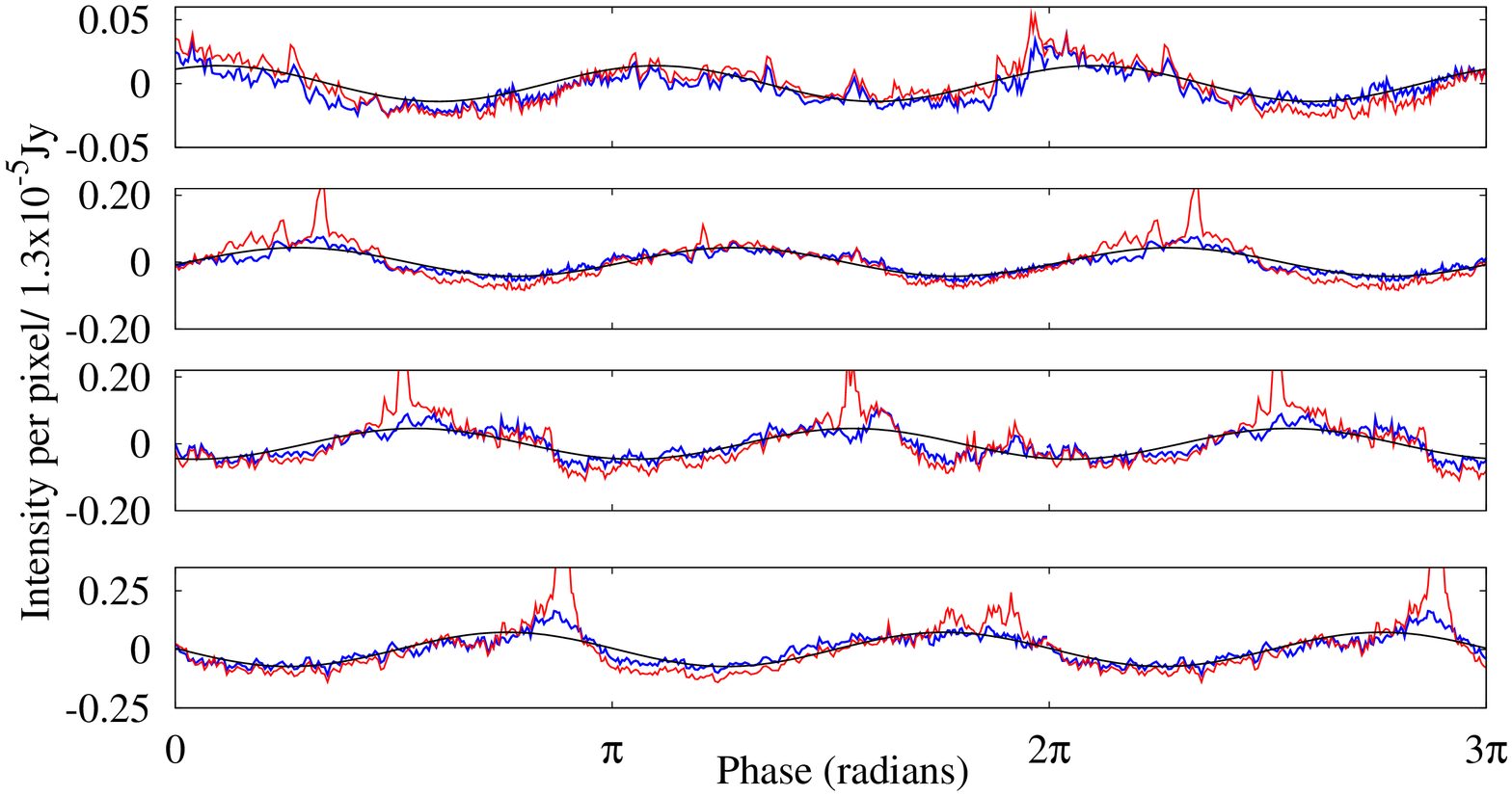}
  \caption{PAH corrected data (blue), and (red) XZAP but non-PAH corrected data at identical SMA values for the 3.6$\mu$m wavelength data. The sine wave fits to the (PAH corrected) data are shown in black. The sharp peaks in excess of the sine wave are stronger in the non-PAH corrected data suggesting that the peaks seen in the PAH-corrected profiles are remnant PAH features.}
  \label{figphase1}
\end{figure}

The phase of the stellar spiral wave can be identified from the peak in the mass surface density on each spiral arm. The variation of this phase with radius reveals useful information about the pitch angle and behaviour of the spiral. The amplitude of the stellar wave is at most $\sim$30 per cent of the mass of the disc and so the signal tends to get hidden in the exponential profile if the total mass surface density is used. Hence, the easiest way to extract the phase information is to exclude the axisymmetric components by using the residual surface density maps. The sine waves fitted to the azimuthal profiles, as described in section \ref{Results}, provide one easy way of measuring the phase (if it is assumed that the spiral structure can be described by a perfect sine wave). The phase of the maximum is obtained automatically from the fit, and a sample of fitted profiles are shown in Fig. \ref{figphase1}. As discussed in section \ref{accuracy}, uncertainties still exist in the residual mass maps, and so a suite of complementary techniques been developed to identify the position of the potential minimum (or maximum in the mass) at each radius sampled. The method of picking the density maximum by fitting a sine wave was subject to a simple check by then examining each of the fits by eye, looking for approximately sinusoidal shape. For each profile the phases of the density maxima were identified (one point per arm). A final approach was to use an automatic routine to pick the two highest peak positions (assuming the arms are $\sim$180$^\circ$ apart). This method has the highest susceptibility of the three to bias from contamination by young bright objects (which will tend to create narrow, bright features). The automatic peak-finding method can be compared to the sine fits and fits by eye in Figs. \ref{figphase2} and \ref{figphase4} where it can be seen that, even after removing as much contamination as possible from the mass surface density maps, the peak finding method picks up traces of the spurs which are clearly seen in the 8$\mu$m image.

In order to justify the fitting of the spiral pattern with just an m=2 sine wave, the power in other Fourier components was analysed. Over the region of interest ($\sim$300-600 arcsec), the power in the next highest component, m=4, averaged only 6 per cent of the power in the m=2 components. Furthermore, examination of the fits demonstrated that much of this power was contributing to the features identified as remnant PAH emission: Fourier analysis of the non-PAH corrected azimuthal profiles showed that, on average, there was 40 per cent more power in the m=4 component than the PAH-corrected profiles (and $\sim$30 per cent more in the m=6 component, the next highest increase). This can be seen qualitatively in  Fig. \ref{figphase1}, where the effect of not removing the PAH component from the 3.6$\mu$m images is clearly shown. The figure shows profiles of intensity vs azimuth (plotted over a 3$\pi$ range), with and without PAH corrections. It is clear that the deviations from a underlying sine wave are much more pronounced in the non-PAH corrected data; it is likely that the remaining non-sine wave features in the azimuthal profiles are a result of not fully removing the PAH components.

\begin{figure}  \centering
\includegraphics[width=83mm]{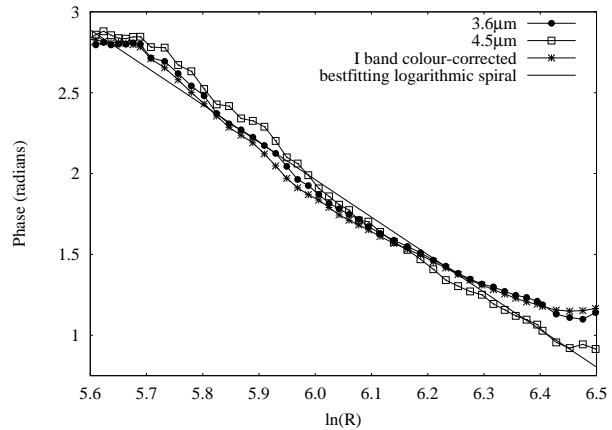}
  \caption{The phase of the stellar spiral structure, as measured by fitting sine waves to the azimuthal profiles. The phase can be fitted with a logarithmic spiral between 5.7$\le$ln(R)$\le$6.4, as shown.}
  \label{figphase3}
\end{figure}

It can be seen from Fig. \ref{figphase4} that although there is scatter in the phase, there is no systematic variation between the two arms, confirming that the arms are symmetric and separated by 180$^\circ$. The lower plot in Fig. \ref{figphase4} illustrates the greater scatter in the phase determined from the peaks in intensity, particularly in the inner parts of the spiral (and is consistent with the presence of remnant PAH emission in the residual images). As already mentioned, this method has the greatest susceptibility to contamination, and although the data are shown for comparison they are not used to calculate the pitch angle or the offset from the peaks in the 8$\mu$m emission. It can also be seen that the phase only shows approximately logarithmic spiral behaviour between 5.7$\le$ln(R)$\le$6.4, corresponding to a range of 300$\le$R$\le$600 arcsec; the inner extent of the spiral wave must clearly lie at R $\sim$300 arcsec. The 600 arcsec cutoff could mark the true extent of the spiral wave in the stellar population, but relatively low signal-to-noise by R $\sim$700 arcsec might also play a part in the loss of log-spiral-like behaviour (see Fig. \ref{figamp1}). Using phase measurements from the sine wave and phase fits by eye the pitch angle can be determined, and is found to be 23$^{o}$. If the sine wave fits alone are considered (Fig.\ref{figphase3}), it can be seen that a steeper gradient (smaller pitch angle) can be fitted to the 4.5$\mu$m than to the 3.6$\mu$m or I band colour-corrected data. The deviations from the straight line (log spiral) behaviour are not insignificant; all three wavelengths show a tendency for phase to increase faster with radius at smaller radii than at the outer extent of the spiral. This causes deviations of 5 and 9$^\circ$ from the phase predicted by the best-fitting gradient for the 3.6$\mu$m at the inner and outer extents of the spiral respectively (and similar for the colour-corrected I band), and 11$^\circ$ at the inner extent of the spiral for the 4.5$\mu$m data. The reason for such close agreement between two of the methods and not the third is not clear, but the difference in pitch angle obtained ($\sim$3$^\circ$) is probably a good indicator of the uncertainties in the method.

\subsection{Phase of the 8$\mu$m spiral structure.}\label{gasphase}

\begin{figure}   \centering
  \includegraphics[width=83mm]{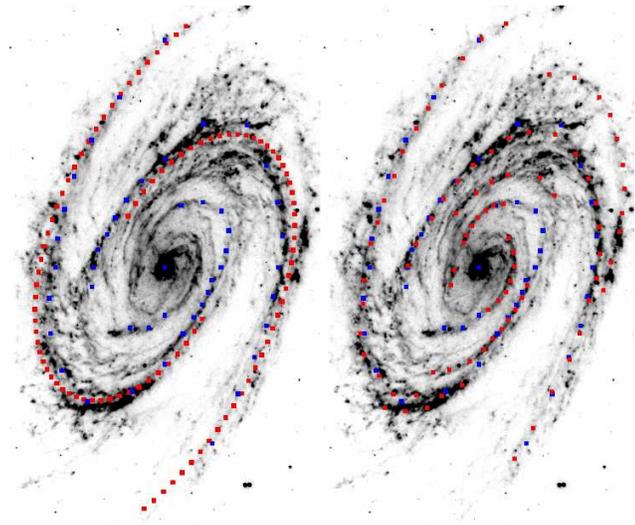}
  \caption{Left; the 8$\mu$m image with (blue) the positions of the peaks in emission as identified by eye from radial cuts, and the positions of the spiral fit to the peaks (red). Right; positions of the peaks in emission as identified by eye (red) and from the radial cuts (blue).}
  \label{figgasphase1}
\end{figure}

The position of the peaks in the 8$\mu$m emission as a function of radius was identified initially by eye from radial cut data (as shown in Fig. \ref{figrad1}). The data were plotted with phase as a function of ln(R), and a logarithmic spiral was assumed, allowing a best-fitting gradient to be determined. This spiral was then overlaid on the 8$\mu$m image, and (small) adjustments to the gradient were allowed until the match between the logarithmic spiral and the image was maximized. The final fit has a pitch angle of 14$^{o}$, shown in Fig. \ref{figgasphase1}.

\begin{figure}   \centering
  \includegraphics[width=83mm]{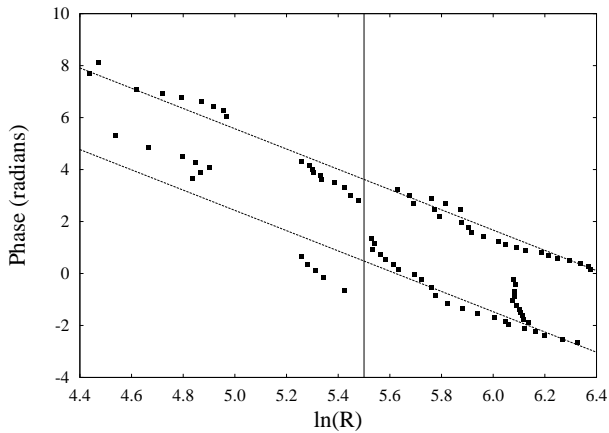}
  \caption{The positions of the peaks of the 8$\mu$m spirals identified directly from the image (deprojected to face-on orientation) with the  best-fitting logarithmic spiral plotted for comparison. The vertical line at ln(R) $\sim$5.5 marks the approximate position of the inner Lindblad resonance; although spiral structure in the gas can be traced well inside this radius the density wave only exists at radii greater than the ILR.}
  \label{figgasphase5}
\end{figure}

The positions of the 8$\mu$m peaks can also be identified directly from the 8$\mu$m image, by using TVMARK in IRAF. The positions identified through this method are shown in Fig. \ref{figgasphase1}. The data are also shown in Fig. \ref{figgasphase5},  viewed as $\phi$ vs ln(R). Interestingly the inner Lindblad resonance (discussed in section \ref{offsetdiscuss}) seems to lie approximately at the radius of an abrupt phase shift in the data.

\subsection{Offsets between the stellar spiral and 8$\mu$m emission.}\label{offset}

With well defined phases for the 8$\mu$m emission and stellar spiral as a function of radius it is possible to look for an offset between the two. When Figs. \ref{figphase2} and \ref{figgasphase1} are compared it becomes apparent that there is a larger offset between the peaks of the best-fitting sine waves and the peaks in the 8$\mu$m emission than between the combined estimates for the density maximum (not exclusively estimated from the sine wave fits) and the 8$\mu$m. This is not surprising since, in section  \ref{massphase}, it was shown that the sine wave fitting was the least susceptible to contamination from remnant PAH features which tend to be on or very near the peak of the 8$\mu$m emission.

\begin{figure*}   \centering
  \vbox to200mm{\vfil 
  \includegraphics[width=180mm]{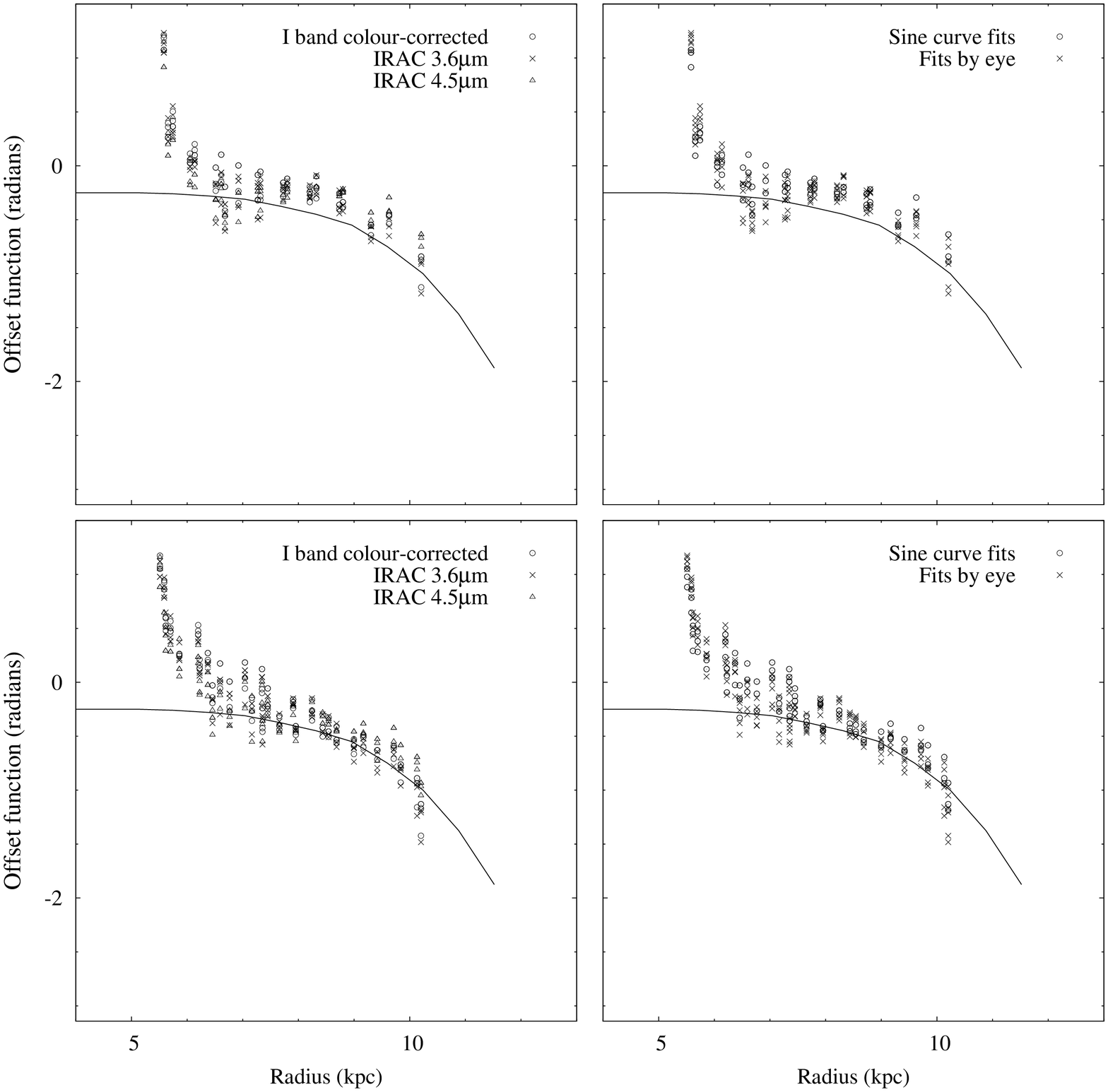}
  \caption{The offset function between the 8$\mu$m emission (predicted to trace the gas shock) and density maximum in the stellar spiral as a function of radius (note, the offset function is twice the actual angular offset). Plots on the top line use the 8$\mu$m phase identified from radial cuts, whereas plots on the bottom use the 8$\mu$m phase determined by eye from the 8$\mu$m image. In each plot the offset function prediction from \protect\cite{2004MNRAS.349..909G} is shown as a solid black line; corotation occurs when the offset function reaches -$\pi$/2. Note that the gas shock is predicted to lie upstream of the density wave as corotation is approached (as is observed).}
  \label{figoffset3}
  \vfil}
\end{figure*}

Fig. \ref{figoffset3} quantifies the angular offset for the different methods used to identify the phase of the stellar spiral and 8$\mu$m emission. The offset is plotted using the offset function, $\Phi$, which for consistency with \cite{2004MNRAS.349..909G} (G\&C) is defined as 2($\phi_{8{\mu}m}$ - $\phi_{stellar-arm})$. The calculation was performed using the data, not the logarithmic spiral fits; the only data excluded are those obtained from the (stellar spiral) phase peak finding method, which, as can be seen in Fig. \ref{figphase2}, often detects spurs rather than the true spiral structure. The trend is for the offset to increase with radius which is in agreement with predictions in G\&C. The model closest to M81 from G\&C is plotted for comparison with this data; corotation is predicted to lie where the offset function reaches -$\pi$.

\section{Discussion.}\label{discuss}

\subsection{Amplitudes.}\label{amplitudediscuss}

\begin{figure}   \centering
  \includegraphics[width=83mm]{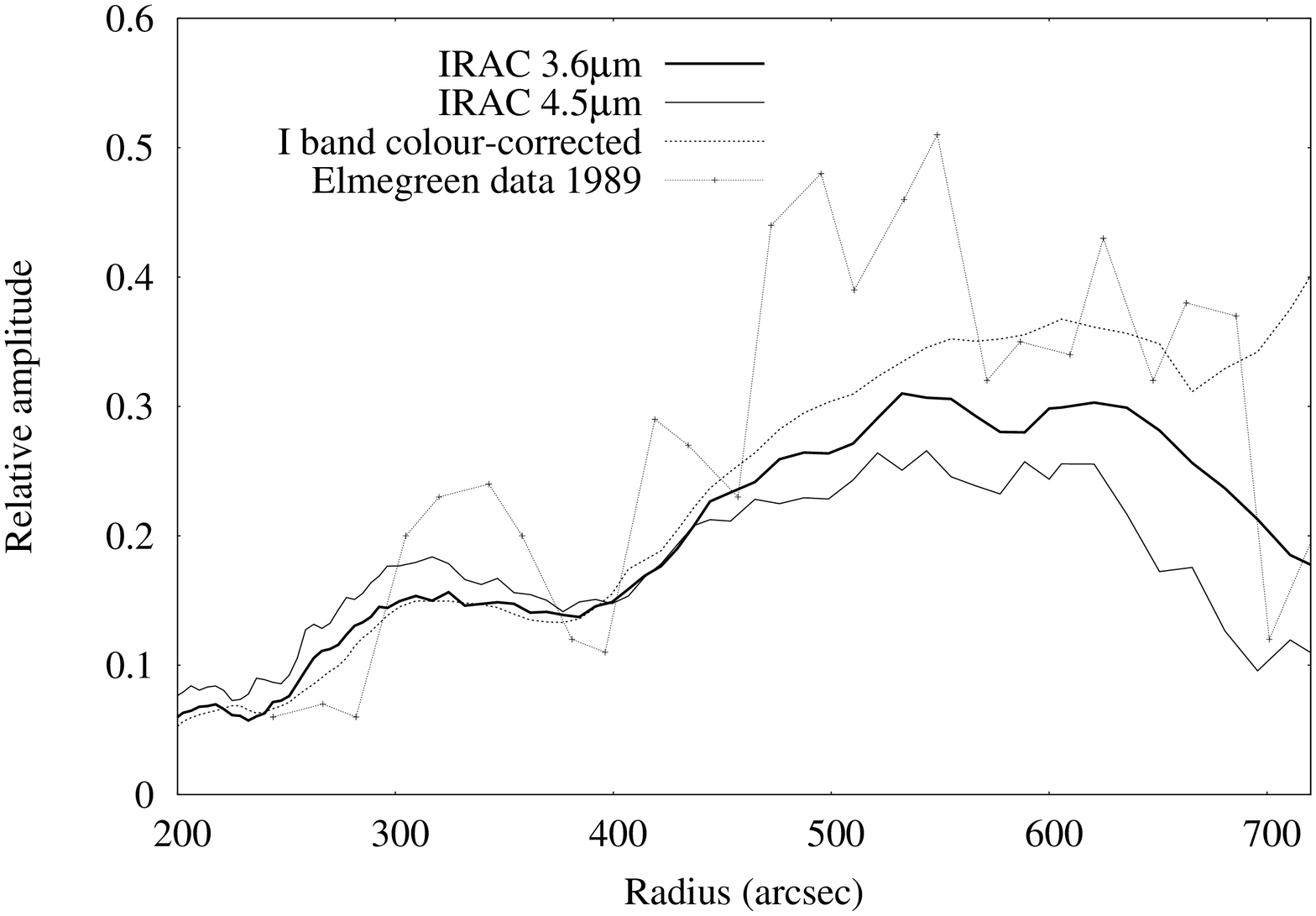}
  \caption{Relative amplitude of the spiral structure determined for the IRAC and \textit{I} band data (as in Fig. \ref{figamp1}) compared to the relative amplitude found by \protect\cite{1989ApJ...343..602E}.}
  \label{figamp2}
\end{figure}

The radial dependence of arm amplitude has been examined in the past, (e.g. \cite{1989ApJ...343..602E}), using the \textit{I} band; these data are also plotted in Fig. \ref{figamp2}.  The results show that both the IRAC and colour-corrected \textit{I} band data display a smoother trend with radius than the Elmegreen data, and the relative amplitude is generally smaller. This may be explained due to contamination effects from young stars in the Elmegreen data (there was no equivalent process to colour correction or PAH removal carried out). It is also worth noting that the relative amplitudes are defined differently; in this paper relative amplitude (in Fourier components) is F$_2$/F$_0$, theirs (F$_2$+F$_6$)/(F$_0$+F$_4$+F$_8$). However, components beyond m = 2 are normally small so to first order these are unlikely to result in large differences. Other authors who have investigated the amplitude of grand design spiral waves include \cite{1976ApJS...31..313S}, who finds the relative amplitude of the spiral arms in the \textit{O} band to vary between $\sim$0.1-0.5. For comparison, the relative amplitudes of the spiral structure in other grand design spirals have been determined; \cite{1989ApJ...343..602E} include M51 and M100 in their study and find that the relative amplitude in M100 varies between 0.1-0.5. The arms in M51 appear to be somewhat stronger, varying between $\sim$0.2-0.8 (in both galaxies the overall trend is for the relative amplitude to increase with radius, as with M81). Schweizer finds the relative amplitude in M51 to increase from 0.2-0.8. \cite{1993ApJ...418..123R} have also studied M51 with \textit{K$_{s}$} band data and find a relative amplitude that varies between $\sim$0.1-0.5, again obeying the general trend to increase with radius. Finally, \cite{1990ApJ...355...52E} use optical data to constrain the spiral arms in NGC 1566, and find a density wave with an average relative amplitude of $\sim$0.4. Values for relative amplitude used in simulations of M81 have tended to be smaller than those found observationally; \cite{1980A&A....88..149V} used models where the relative amplitude ('forcing') varied between 0.05-0.1. \cite{2004MNRAS.349..909G} use a model for M81 with relative amplitude $\sim$0.03, and investigate amplitudes no greater than 0.1 in the non-M81-specific models.

It should be noted that the large error bars associated with the relative amplitude (as shown in Fig. \ref{figamp1}) are a conservative estimate of the error in measuring the amplitude of the wave. Given that the random scatter in the amplitudes for each wavelength is much smaller, the errors given should be considered to be an estimate of the systematic error associated with the measurement. 

\subsection{Offsets.}\label{offsetdiscuss}

\cite{2004MNRAS.349..909G} (G\&C) used semi-analytic theory supplemented by hydrodynamic simulations to compute the steady state response of isothermal gas to a rigidly rotating spiral mode. For given galaxy input parameters (i.e. rotation curve, relative amplitude of mode, azimuthal wave number) it is found that, inward of corotation, the shock in the gas moves steadily upstream with respect to the spiral arms in the stellar potential as radius is increased. As corotation is approached, the angular offset increases rapidly. In principle, therefore, offset data should readily indicate the corotation radius. In practice, however, the weakening of the shock as corotation is approached means that it is only feasible to obtain observational constraints on offsets at radii well within corotation and G\&C therefore present estimates of the accuracy with which the corotation radius can be determined as a function of the completeness of the data available. As a general guideline, G\&C suggest that $R_{co}$ can be determined to within 25 per cent if the angular offsets observed give an offset function range exceeding $\sim \pi/4$ (for a two armed spiral, offset function is twice the angular offset). 

If the observed spiral structure in M81 is assumed to be a long-lived density wave which maintains a well defined, constant pattern speed over a number of galactic rotations, it is possible to compare our results to those of G\&C. When the data in Fig. \ref{figoffset3} are examined, corotation appears to lie at $\sim$13kpc. This result has, for the moment, assumed that any discrepancies between the model parameters in G\&C compared to those determined in this paper can be ignored. This is largely justifiable due to small differences in the relevant parameters, however the difference between the relative amplitude of the spiral perturbations in the model used by G\&C and our data are rather larger. To compare the relative amplitude quoted in this work to the 'forcing' used to drive the simulated gas shocks in G\&C it is necessary to include the halo contribution to the axisymmetric mass. \cite{blok} have used SINGS and THINGS (The HI Nearby Galaxies Survey) data to constrain the halo mass, and in the case of M81 find that the fraction of mass in the halo to that in the disc+bulge increases from $\sim$0.12-0.25 in the range 300$\le$R$\le$600 arcsec. Using this additional contribution to the axisymmetric components, the relative amplitude of the density wave to total axisymmetric mass (equivalent to the forcing employed by G\&C) then varies between 0.12-0.26. The standard value for the relative amplitude used by G\&C was 0.05 at 8.5kpc. In contrast, we find the relative amplitude to be closer to 0.225 at 8.5kpc \footnote{Note that there is some evidence from G\&C that the shock moves further downstream at small radii in the case of larger density wave amplitudes. This may explain the fact that the pronounced shift to positive offsets at the smallest radii in the M81 data is not reflected in the model.}. As noted by G\&C, for a given radius (and fixed corotation), increasing wave amplitude causes the offset to decrease rather than increase. Another factor that affects the offset function in the models is the assumed sound speed in the gas: G\&C found that at fixed R/R$_{CR}$ the shock moves further upstream in the case of warmer gas (see also \cite{2007MNRAS.376.1747D} who found that the shock was downstream even at relatively large R/R$_{CR}$ for their very cold gas, although it should be noted that this result is also found in warm gas by G\&C for the m=4 case - inapplicable to M81 - that is treated by \citeauthor{2007MNRAS.376.1747D}). The model of G\&C plotted in Fig. \ref{figoffset3} assumes a sound speed of $10$ km s$^{-1}$.  We thus deduce that either the use of colder gas or the use of a larger spiral arm amplitude would cause the fraction R/R$_{CR}$ at a given radius to be larger than predicted by the model shown in Fig. \ref{figoffset3}. In addition, as shown in Fig. \ref{figamp1}, the relative amplitude of the arms appears to increase with radius, and G\&C note that a systematic variation in wave amplitude over the disc may lead to an inaccurate estimate for corotation. Taking all the uncertainties caused by relative amplitude into account, corotation is estimated at  $\sim$12kpc $\pm$3kpc (estimating an error of 25 per cent according to the guidelines given by G\&C).

\begin{table}
  \centering
  \begin{tabular}{|l|l|} 
    \hline
    Author & R$_{CR}$ (kpc) \\
    \hline
    Elmegreen et al. (1989) & 9 \\ 
    Gottesman \& Weliachew (1975) & 11.7-12.6 \\ 
    Lowe et al. (1994) & 10  \\ 
    Roberts et al. (1975) & 11 \\ 
    Rots (1975) & 11 \\ 
    Sakhibov \& Smirnov (1987) & $>$12 \\ 
    Shu et al. (1971) & 15 \\ 
    Tamburro et al (2008) & 8.7 $\pm$ 4.7\\
    Visser (1980) & 12 \\ 
    Westpfahl (1998) & 9 \\  
    This work & 12 $\pm$ 3 \\
    \hline
  \end {tabular}
  \caption{Values for R$_{CR}$ found previously. The methods used are described further in the text. All values are scaled to the current M81-distance estimate of 3.6Mpc. Errors are shown when stated by the authors, otherwise R$_{CR}$ is given to nearest kpc.}
  \label{tab2}
\end{table}

There have been many estimates made of the value of corotation in M81; a sample are listed in Table \ref{tab2}.  The methods have varied, but brief descriptions of a number are offered below. \cite{1971ApJ...166..465S} uses the outermost HII regions to define the radius of corotation. \cite{1975ApJ...196..381R} use HII measurements from \cite{1959PASP...71..101M} in combination with spiral structure and the extent of the visible disc and, assume that R$_{CR}$ must be approximately coincident with the radial extent of all three. \cite{1975A&A....45...43R} defines a pattern speed  ($\Omega_p$ = 20kms$^{-1}$kpc$^{-1}$) which, when combined with the rotation curve, gives believable estimates for both the ILR and R$_{CR}$. The estimates for the ILR and R$_{CR}$ assume that the density wave must be constrained between the ILR and R$_{CR}$. HI morphology is used to define these inner and outer limits. In contrast, \cite{1975ApJ...195...23G} use HI data to calculate the radial wavelength of the spiral pattern and, together with the epicyclic frequency and mass surface density, use the method of \cite{1969ApJ...155..721L} to calculate $\Omega_p$ and from that R$_{CR}$. \cite{1987SvA....31..132S} use HI velocity data. The axisymmetric rotation curve is modeled iteratively, and non-circular perturbations are used to calculate R$_{CR}$ (the uncertainty in the result is due to the rotation curve being insufficiently defined). \cite{1998ApJS..115..203W} uses the Tremaine \& Weinberg method \citep{1984ApJ...282L...5T}. \cite{1989ApJ...343..602E} use \textit{B} and \textit{I} band observations to identify resonance features (specifically the 4:1 resonance) in the spiral arm amplitudes. \cite{1994ApJ...427..184L} also use data from \cite{1989ApJ...343..602E} to build a model of M81 based on modal theory. \citeauthor{Tamburro} use the offset between the peaks of star formation and cold atomic hydrogen measured via 24$\mu$m and HI emission respectively. The offsets are calculated for a number of radii by the cross-correlation of azimuthal profiles of the two wavelengths, and corotation is expected where this offset falls to zero. Finally, \cite{1980A&A....88..149V} chooses $\Omega_p$ such that the ILR and R$_{CR}$ have believable values for the rotation curve used in his models (such that wave propagation is possible over a suitable radial range). 

\begin{figure}   \centering
  \includegraphics[width=83mm]{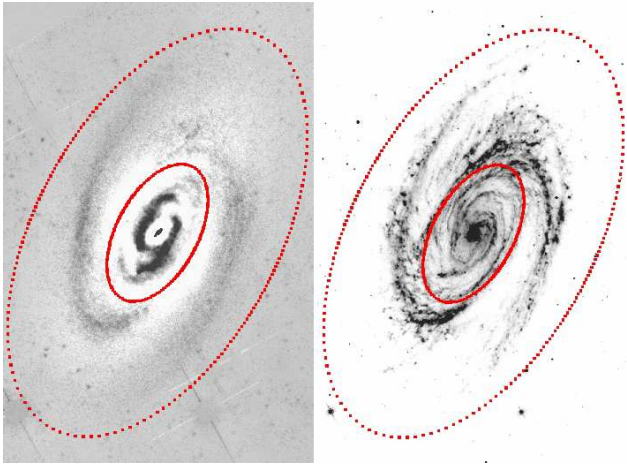}
  \caption{The ILR and corotation circles (12.8kpc and 4.3kpc) indicated on the 3.6$\mu$m residual mass surface density map image (left) and 8$\mu$m image (right).}
  \label{resonances}
\end{figure}

Given the range of values quoted above, R$_{CR}$ = 12-13kpc is not inconsistent. Using this value, combined with the M81 rotation curve (see, e.g. \cite{1999ApJ...523..136S}), the pattern speed is found to be $\sim$17kms$^{-1}$kpc$^{-1}$. The inner Lindblad resonance (ILR) can also be estimated using the rotation curve and $\Omega_p$, and is found to lie at $\sim$4.3kpc. The positions of corotation and the ILR can be seen on Fig. \ref{resonances}. Corotation lies beyond the observable stellar wave and 8$\mu$m spiral structure, which is consistent with the theory that the shocks should weaken as corotation is approached. The ILR is expected to act as a barrier to density wave propagation (\cite{1969ApJ...158..899T, 1971PNAS...68.2095M}), so if the ILR is indeed located at 4.3kpc the dramatic fall in amplitude at this radius can be explained by the presence of the ILR, rather than modal amplitude damping as has been claimed in the past -  e.g. \cite{1989ApJ...343..602E}. Although the ILR should mark the inner edge of (stellar) spiral structure, the inner mini-ring and stubby spiral arms at $\sim$100 arcsec (described in section \ref{overview}) are well inside this radius. The most likely explanation is that these features are driven by the spiral structure in the gas (rather than vice versa), as the mini-ring and spiral arms clearly coincide with structures in the 8micron image - the ILR does not affect the gas, and so spiral structure is able to propagate through to the regions within. However, it is also possible that a weak oval distortion could provide the driving force. It should be noted that, as shown in figure \ref{figgasphase5}, the spiral structure appears to become more filamentary for R$<$R$_{ILR}$, which would seem to support these conclusions.

The agreement between the observed offset and theory is particularly interesting because the models of G\&C assume that the galaxy has had time to reach steady-state behaviour. If the spiral structure is highly transient the offset between density wave and gas shock would not have time to develop, hence, at least in the case of M81 it appears that the spiral arms must be reasonably long-lived, at least for the order of a few dynamical timescales. 

\subsection{Fourier decomposition.}\label{fourierdiscuss}

The lack of power in the non m=2 components of the harmonic decomposition (described in section \ref{massphase}) is initially surprising, particularly the relatively small contributions to m=4. \cite{1989ApJ...338...78B} studied global modes in spiral galaxies and concluded that it is not easy to support single dominant modes of spiral structure in their models (whilst observing that such galaxies do exist), and suggest that two or three important modes are more stable. It is quite possible that the external influences on M81 from M82 and NGC 3077 (see section \ref{interact}) are the cause of the strength of the m=2 mode, although the excitation of modes by an external influence is not necessary for the existence of spiral structure in the modal picture. \citeauthor{1989ApJ...338...78B} also noted that the stability of spiral structure over a number of rotation periods is reliant on the existence of one dominant mode; a larger number of modes will tend to cause the spiral to evolve gradually in a quasi-periodic manner. Thus, the observation of a single dominant mode in M81 is in fact consistent with the inferred long-term stability of the spiral structure from the behaviour of the offsets between the stellar spiral and gas shock.

\subsection{Interactions and tidal driving mechanisms.}\label{interact}

\begin{table}
  \centering
  \begin{tabular}{|l|l|l|l|} 
    \hline 
    Name & R$_{rad}$ (kpc) & R$_{closest}$ (kpc) & V$_{R}$ (kms$^{-1}$)\\ 
    \hline
    M82 & 38.6 & 29.0 & $\sim$250 \\
    NGC 3077 & 48.4 & 18.6 & $\sim$50\\
    \hline
  \end {tabular}
  \caption{R$_{rad}$ is the projected distance to M81 from the companion galaxy. V$_{R}$ is given relative to M81. Distances of closest approach are taken from the simulations of \protect\cite{1999IAUS..186...81Y}. The distance of the M81 group is taken to be 3.6Mpc.}
  \label{tab3}
\end{table}

Simulations by \cite{1999IAUS..186...81Y} based on HI observations demonstrate that M81, M82 and NGC 3077 underwent a three-way interaction approximately 2x10$^{8}$ years ago (nearest approaches of M82 and NGC 3077 to M81 were 2.2 and 2.8x10$^{8}$ years ago respectively). The current (radial) velocities and projected radial distances of M82 and NGC 3077 with respect to M81 are given in Table \ref{tab3}, along with the distances of nearest approach taken from Yun's simulations. It is immediately apparent that NGC 3077 cannot be driving the spiral structure in M81; the relative velocity is too small. 

In contrast, it seems entirely possible that M82 could provide the driving force behind the spiral structure observed in M81. Given the limited information available all calculations must be approximate, but if it is assumed that the pattern speed of the spiral must match the angular speed of the interaction between M82 and M81, the speed and radius of M82 are certainly of the correct order of magnitude; M82 is required to have a velocity relative to M81 of $\sim$500 kms$^{-1}$ at perigalacticon to match the pattern speed of the spiral (the matching of pattern speed and angular interaction is not necessary, but the response of the stellar spiral will be stronger in this case). Further, the timescale of this interaction (a few x10$^{8}$ years), matches the minimum predicted by G\&C that is needed to observe an offset between the density wave and gas shock.

\section{Conclusions.}\label{conclusions}

We have used near infrared and optical images to produce mass surface density maps of M81 using IRAC 3.6 and 4.5$\mu$m data, optical \textit{B}, \textit{V} and \textit{I} band images, and 2MASS K$_{s}$ band data. We have shown that there is an underlying spiral wave in the old stellar population of M81. This spiral structure is not unbroken throughout the galaxy, but can be traced through nearly a full 180$^{o}$ rotation between 300$<$R$<$600 arcsec. The amplitude of the stellar wave is found to rise from $\sim$0.1-0.3 over this range. The amplitude estimates from all three methods agree to within the uncertainties, although systematic differences in the measured relative amplitude appear at larger radii. The pitch angle of the spiral can be determined within this range and is found to be 23 $\pm$ 3$^{o}$.

By using the IRAC 8$\mu$m band as a indicator of the position of shocks in the gas we have measured the phase of the shocks induced by the stellar mass variations. The angular offset between the gas shocks and stellar spiral was used to determine the corotation radius, by using the method put forward by \cite{2004MNRAS.349..909G}, which assumes that the spiral pattern rotates as a rigid body (ie, a density wave). Corotation was determined to lie at 12 $\pm$ 3kpc. The position of the inner Lindblad resonance was extrapolated from the rotation curve and pattern speed, and found to lie at 4.3kpc. These results are consistent with previous estimates, and with the observed morphology of the stellar spiral and gas structure. In addition, the offset between the stellar spiral wave and gas shock suggests that steady-state behaviour must have been reached, meaning that a density wave with a constant pattern speed must have survived for at least a few x10$^{8}$ years. Further numerical simulations are required in order to investigate whether the observed offsets are consistent with the spiral structure originating from a galactic encounter.

The dynamics and projected distances of M81's nearest neighbours, M82 and NGC 3077, are examined, and by matching the pattern speed of the spiral arms with the angular speed of the companion galaxy (around M81), it is shown that the interaction between M81 and M82 could potentially provide the driving force for the density wave. The timescale of this interaction (nearest approach $\sim$2.2x10$^{8}$ years ago) is consistent with the (minimum) time needed to set up a steady state response to the density wave in the gas.

A logical extension to this work is to extend the sample to more galaxies. There are many candidate spiral galaxies in the SINGS survey, and a larger sample will allow a larger range of questions to be addressed, such as investigating further the link between stellar spirals, induced shocks in the interstellar gas, and star formation (as carried out by \cite{2002MNRAS.337.1113S}, for example).

\section{Acknowledgments.}\
Many thanks to Chad Engelbracht for the optical images of M81, and to Mike Regan and Daniela Calzetti for the original IRAC image processing. We are grateful to Erwin de Blok for allowing us access to unpublished results from the THINGS project, and to Mike Regan, Hans-Walter Rix, and Jim Pringle for many helpful suggestions and advice. In addition, we thank the referee, Preben Grosbol, for his comments.

This work makes use of IRAF. IRAF is distributed by the National Optical Astronomy Observatories, which are operated by the Association of Universities for Research in Astronomy, Inc., under cooperative agreement with the National Science Foundation.

\bibliography{bib}

\label{lastpage}

\end{document}